\documentclass[review,12pt]{elsarticle}
\usepackage{lineno,hyperref}
\modulolinenumbers[5]
\usepackage[utf8]{inputenc}
\usepackage{amsmath}
\usepackage{amsfonts}
\usepackage{amssymb}
\usepackage{lmodern}
\usepackage{xfrac}
\usepackage{graphicx}
\usepackage{subfig}
\graphicspath{ {./figures/} }
\usepackage{fullpage}
\usepackage{float}
\usepackage{caption}
\usepackage{siunitx}
\usepackage{bm}
\DeclareMathOperator{\Tr}{Tr}
\usepackage{xcolor}

\biboptions{sort&compress}

\usepackage{lipsum}
\makeatletter
\def\ps@pprintTitle{%
 \let\@oddhead\@empty
 \let\@evenhead\@empty
 \def\@oddfoot{}%
 \let\@evenfoot\@oddfoot}
\makeatother

\begin{document}

\begin{frontmatter}

\title{Simulation of temperature, stress and microstructure fields during laser deposition of Ti-6Al-4V}

\author[mymainaddress]{Supriyo Ghosh\corref{mycorrespondingauthor}}
\cortext[mycorrespondingauthor]{Corresponding author.}
\author[mymainaddress]{Kevin McReynolds}
\author[mymainaddress]{Jonathan~E.~Guyer}
\author[mymainaddress]{Dilip Banerjee}
\address[mymainaddress]{Materials Science and Engineering Division, National Institute of Standards and Technology, Gaithersburg, MD 20899, USA}

%


\begin{abstract}
We study the evolution of prior columnar $\beta$ phase, interface $L$ phase, and $\alpha$ phase during directional solidification of a Ti-6Al-4V melt pool. Finite element simulations estimate the solidification temperature and velocity fields in the melt pool and analyze the stress field and thermal distortions in the solidified part during the laser powder bed fusion process. A phase-field model uses the temperature and velocity fields to predict the formation of columnar prior-$\beta$(Ti) phase. During the solidification of $\beta$ phase from an undercooled liquid, the residual liquid below the solidus temperature within the $\beta$ columns results in $\alpha$ phase. The finite element simulated stress and strain fields are correlated with the length scales and volume fractions of the microstructure fields. Finally, the coalescence behavior of the $\beta$(Ti) cells during solidification is illustrated. The above analyses are important as they can be used for proactive control of the subsequent modeling of the heat treatment processes.
\end{abstract}


\end{frontmatter}
\section{Introduction}
Ti-6Al-4V is an important engineering alloy due to its excellent strength-to-weight ratio, bio-compatibility and corrosion resistance and is therefore widely used in aerospace, biomedical and marine applications~\cite{Herzog2016,Korner2016,Tan2015}. During the laser powder bed fusion (LPBF) additive manufacturing (AM) process, parts are built layer-by-layer by rastering the laser across the alloy powder followed by melting and solidification processes. Solidification occurs rapidly in the trailing edge of the molten pool, resulting in columnar and/or equiaxed microstructures which determine the properties of as-built Ti-6Al-4V~\cite{Herzog2016,Korner2016}. The solidification conditions, namely the temperature gradient and velocity fields in the molten pool are often estimated using finite element analysis (FEA)~\cite{Suresh2016,Trevor2017,ghosh2018_single}. FEA simulations can also predict the stress and strain fields within the solidified melt pool, which are often correlated with the length scale of microstructure phase fields~\cite{Tan2015,Raeisinia2008}. A macroscopic stress analysis of the entire workpiece can predict the residual stress distribution and distortions during and after the laser deposition process. During the solidification and subsequent cooling period, plastic strains result in the deposited part both due to temperature variations and constraints of clamping forces.

Ti-6Al-4V is an $\alpha$-$\beta$ alloy. Al stabilizes $\alpha$ which imparts solid solution strengthening, and V stabilizes $\beta$ which improves ductility and fatigue properties. Liquid transforms to body-centered-cubic prior-$\beta$ phase after solidification. $\beta$ transforms to hexagonal-closed-packed $\alpha$ phase below the $\beta$ transus temperature. During $\beta \leftrightarrow \alpha$ transformation, a small amount of face-centered-cubic $L$ phase forms at the $\alpha$/$\beta$ interface during the intermediate stages which reduces tensile and fatigue strengths of the material. The size, morphology and distribution of $\beta$, $\alpha$ and $L$ phases within the as-solidified Ti-6Al-4V are therefore important to consider and are estimated using a phase-field model. Post-deposition heat treatments are often used to alter properties of $\alpha$, while it does not have a large effect on prior-$\beta$ phase. Therefore, the present work focuses on modeling the formation of prior-$\beta$(Ti) phase through a phase-field approach using FEA predicted thermal field and macroscopic stress analysis during laser deposition of Ti-6Al-4V. 

There have been several studies regarding the LPBF modeling of Ti-6Al-4V (for recent reviews, refer to Refs.~\cite{Herzog2016,Gorsse2017,Luo2018}). Karayagiz et al.~\cite{kubra2018} reported the numerical and experimental analyses of temperature distribution during LPBF of Ti-6Al-4V. Sahoo et al.~\cite{Chou2016} estimated the thermal history in a Ti-6Al-4V molten pool using a heat transfer FEA model and simulated the dendritic solidification microstructures using a phase-field model. However, the authors in Ref.~\cite{Chou2016} did not determine the mechanical history of the as-solidified Ti-6Al-4V part and hence no microstructure-properties correlation were reported. Luo et al.~\cite{Luo2018} surveyed the FEA of temperature and thermal stress fields that evolve during LPBF of Ti-6Al-4V, but the microstructures for the corresponding thermal history were not simulated. Lu et al.~\cite{Lu2018} modeled only the distortion and residual stresses in a Ti-6Al-4V part during a laser solid forming process. Trends in solidification grain size and morphology for any practical range of AM process variables of Ti-6Al-4V has been discussed by Gockel et al.~\cite{Gockel2017}. However, there has been no comprehensive study of coupled/uncoupled thermo-mechanical and microstructure analyses of laser powder bed fusion of Ti-6Al-4V. Several process deficiencies such as residual stress induced distortion that arise due to large temperature gradients during heating and cooling cycles in Ti-6Al-4V are rarely studied~\cite{Dunbar2016}. Therefore, numerical analyses of temperature, stress, and microstructure fields that evolve in an as-built Ti-6Al-4V part are essential to study the relationship between LPBF parameters, microstructures and mechanical performance of the final part, saving significant time and cost when compared with conducting numerous physical experiments.

\section{Powder bed fusion simulation}
The macroscopic FEA simulations address both the heat flow modeling during the laser deposition process and the stress analysis of the part. A sequential analysis procedure (weak coupling) has been adopted here where a finite element heat transfer analysis is conducted first followed by a finite element stress analysis by including the temperature distribution in the model obtained from the heat transfer analysis. It should be noted that a fully coupled heat transfer-stress analysis is computationally prohibitive given the size of the present FEA model and number of computational steps involved in this study.

\subsection{FEA heat transfer analysis}
\begin{figure}
\begin{center}
\includegraphics[scale=0.35]{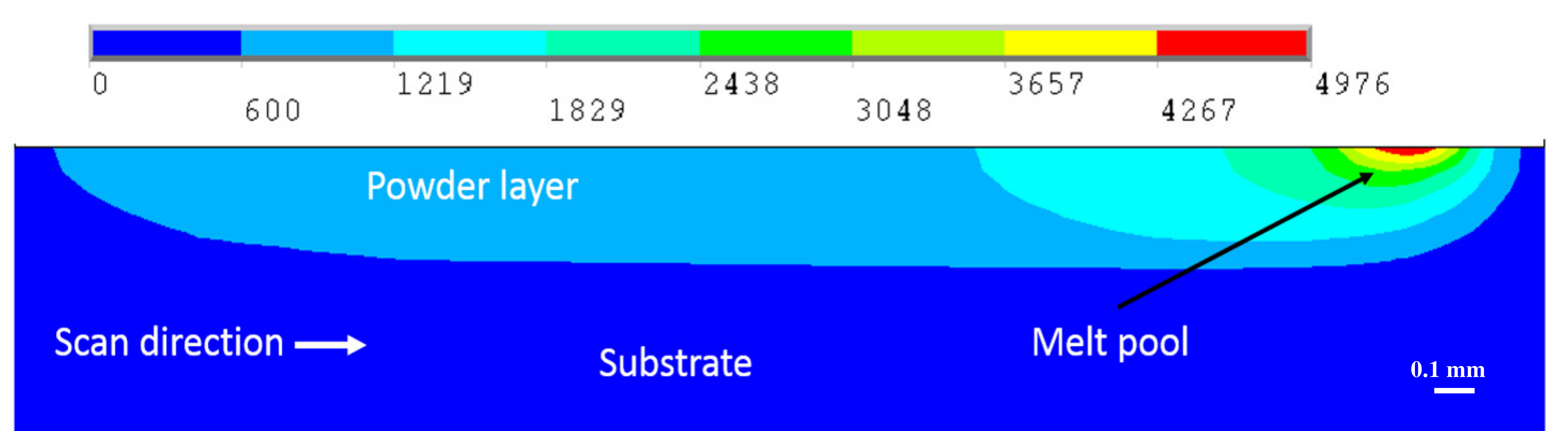}
\caption{A 2D slice plot of temperature distribution in the part after the laser has traversed 80\% of the length for the case of the 20 W laser. The temperatures are in degrees Celsius. The temperatures reported in this study reasonably match with the temperature values reported in Ref.~\cite{Fu2014}. Laser scan direction is $y$ and the (vertical) build direction is $z$. The size of the computational domain: 2.5 mm in length ($y$), 1 mm in width ($x$), and 0.5 mm in height ($z$). The figure is a slice through the computational domain in the longitudinal direction $y$.}\label{figure_fea}
\end{center}
\end{figure}

The analysis is conducted on a rectangular parallelepiped specimen as shown in the elevation view in Fig.~\ref{figure_fea}. The geometry consists of a base part of a Ti-6Al-4V alloy with 2.5 mm length, 1 mm width, and 0.47 mm height. A layer of Ti-6Al-4V powder is laid over this base specimen with a height of \SI{30}{\micro\meter}. The powder bed is discretized by 5 elements along the depth. A laser of \SI{50}{\micro\meter} diameter is focused on the central longitudinal line ($x$ = 0.25 mm, $y$ = 0 mm to 2.5 mm, $z$ = 0) on the top surface and is moved at the rate of 800 mm s$^{-1}$ to complete one end to end run along the length. A total of 1187500 ANSYS\footnote{Certain commercial equipment, instruments, or materials are identified in this paper in order to specify the experimental procedure adequately. Such identification is not intended to imply recommendation or endorsement by NIST, nor is it intended to imply that the materials or equipment identified are necessarily the best available for the purpose.} SOLID 70 elements~\cite{Ansys} describe the base plate region and 250000 ANSYS SOLID 70 elements describe the powder region for the thermal analysis. Element size is much finer in the powder region and in the top region of the substrate. The temperature-dependent thermophysical properties of the Ti-6Al-4V alloy (both solid/liquid material and the powder) needed for the heat flow modeling are taken from Refs.~\cite{Fu2014,boivineau2006,verhaeghe2009,asm_handbook} and are given in Table~\ref{table_conductivity}, Table~\ref{table_conductivity_powder}, Table~\ref{table_specific_heat}, and Table ~\ref{table_others}.
\begin{table}[htbp]
\begin{center}
\caption{Thermal conductivity ($k$) of Ti-6Al-4V (solid) values used in FEA simulation, after~\cite{Fu2014,boivineau2006,verhaeghe2009,asm_handbook}.}\label{table_conductivity}
  \begin{tabular}{| c | c |}
    \hline
    Temperature ($^\circ$C) & $k$ (W m$^{-1}$ K$^{-1}$) \\ \hline
    20 & 26.85 \\ \hline
     100 & 8.15 \\ \hline
     200 & 9.44 \\ \hline
     500 & 13.32 \\ \hline
     876.85 & 18.2 \\ \hline
    1000 & 19.79 \\ \hline
    1500 & 26.26 \\ \hline
    1665 & 28.27 \\ \hline
    2126 & 37.00 \\ \hline
    2426 & 42.00 \\ \hline
  \end{tabular}
  \end{center}
  \end{table}
  \begin{table}[htbp]
\begin{center}
\caption{Thermal conductivity ($k$) of Ti-6Al-4V powder values used in FEA simulation, after~\cite{Fu2014,boivineau2006,verhaeghe2009,asm_handbook}.}\label{table_conductivity_powder}
  \begin{tabular}{| c | c |}
    \hline
    Temperature ($^\circ$C) & $k$ (W m$^{-1}$ K$^{-1}$) \\ \hline
    20 & 0.2 \\ \hline
    1605 & 19.4 \\ \hline
    1665 & 28.3 \\ \hline
  \end{tabular}
  \end{center}
  \end{table}
  \begin{table}[htbp]
\begin{center}
\caption{Ti-6Al-4V solid and powder specific heat ($C_p$) values used in FEA simulation, after~\cite{Fu2014,boivineau2006,verhaeghe2009,asm_handbook}.}\label{table_specific_heat}
  \begin{tabular}{| c | c |}
    \hline
    Temperature ($^\circ$C) & $C_p$ (J kg$^{-1}$ K$^{-1}$) \\ \hline
20 & 580  \\ \hline
205 & 610  \\ \hline
425 & 670  \\ \hline
650 & 760  \\ \hline
870 & 930  \\ \hline
1000 & 936  \\ \hline
1200 & 1016  \\ \hline
1400 & 1095  \\ \hline
1665 & 1126  \\ \hline
  \end{tabular}
  \end{center}
  \end{table}
\begin{table}[htbp]
\begin{center}
\caption{Ti-6Al-4V solid and powder properties used in FEA simulation, after~\cite{Fu2014,asm_handbook}.}\label{table_others}
  \begin{tabular}{| c | c |}
    \hline
    Density (kg m$^{-3}$) & 4428 \\ \hline
 	Latent heat (J kg$^{-1}$) & 365000 \\ \hline
 	 Solidus temperature ($^\circ$C)  & 1605 \\ \hline
 	 Liquidus temperature ($^\circ$C) & 1655 \\ \hline
  \end{tabular}
  \end{center}
  \end{table}\\
The heat transfer analysis~\cite{Ansys} solves the equation for the conservation of energy given by
\begin{equation}\label{eq_fea1}
\rho C_p \left(\frac{\partial T}{\partial t} + \vec{v} \cdot \nabla T \right) + \nabla \cdot \vec{q} = \dot{Q},
\end{equation}
where $\rho$ is the density, $C_p$ is the specific heat capacity, $T(x,y,z,t)$ is the temperature, $t$ is the time, $\vec{v}$ is the velocity vector, $\vec{q}$ is the heat flux vector, and $\dot{Q}$ is the heat source term.

The initial condition assumes a uniform preheat temperature of 79 $^\circ$C in the powder and the substrate at time $t = 0$. Equation~(\ref{eq_fea1}) is solved as a transient analysis (full Newton-Raphson with line search along with the use of an iterative solver) with the following boundary condition: all exposed surface (except those that have laser, see below) are assigned a Newtonian convective boundary condition with a film coefficient of 30 W m$^{-2}$ K$^{-1}$ and ambient temperature of 20 $^\circ$C. No lumped mass matrix approximation was used. The phase change phenomenon is addressed using temperature dependent enthalpy data entered for the powder and the homogeneous base material. It may be noted here that the phase change should be typically handled through the source term in Eq.~(\ref{eq_fea1}) when a macro-microscopic coupling is desired. Since a phase-field approach is used in this study for microstructure calculation, it will be computationally prohibitive to provide this linkage for a part consisting of over 1 million elements. Therefore, the phase change is handled through temperature dependent enthalpy data. The energy equation~\ref{eq_fea1} is solved using a modified specific heat approach, as given in Ref.~\cite{Rappaz1988}. The model has been constructed in ANSYS 17.2 using ANSYS Parametric Design Language (APDL)~\cite{Thompson2017}. An implicit time integration scheme has been used in the FEA simulation. A local coordinate system is used to define the center of the laser, which is updated after each simulation (at a particular laser position) as the laser traverses along the length. This allows for easy assignment of the laser heat flux boundary condition. The laser is a continuous Nd:YAG laser with wavelength of \SI{1.06}{\micro\meter}. The absorption coefficient (= 0.77) is assumed to be the same as that for pure Ti powder~\cite{Tolochko2000}. A Gaussian distribution of the heat flux is assumed following~\cite{Fu2014}. Five different laser powers were considered: 20 W, 40 W, 60 W, 80 W, and 100 W. The goal here is to determine the effect of laser power on temperature gradient, microstructure, and the resulting stress distribution in the deposited part. The elements belonging to the powder region and the substrate are assigned different thermophysical properties and are assigned different material IDs at the beginning of the simulation. If the computed nodal temperatures of all nodes of an element reach the solidus temperature ( = 1878 K), then that element is assigned an ID which is the same as that of the substrate (solid) material. Subsequently, this element retains this same material ID as the laser moves away to a different spot. Common challenges associated with element birth/death are not present here as the current model considers only one layer of powder on top of the substrate.

\subsection{FEA structural analysis}
Following thermal analysis, stress analysis is conducted in ANSYS using the temperature distribution obtained from the heat transfer analysis. A stress analysis provides information on deformation and residual stresses in the deposited part. The governing nonlinear stress equation is given as:

\begin{equation}\label{eq_fea2}
\nabla\cdot\bm{\sigma} + \bm{f} = 0.
\end{equation}
where $\bm{\sigma}$ is the stress tensor and $\bm{f}$ is internal force. Considering an elasto-plastic behavior for the material, strain and stress tensors can be described by the equation
\begin{equation}\label{eq_fea3}
\bm{\sigma} = \bm{C} \bm{\epsilon},
\end{equation}
where $\bm{C}$ is the fourth-order material stiffness tensor and the total strain tensor $\bm{\epsilon}$ is described by the sum of elastic strain $\bm{\epsilon}_e$, the thermal strain $\bm{\epsilon}_{t}$, and the plastic strain $\bm{\epsilon}_p$ as shown in Eq.~(\ref{eq_fea4}). The $\bm{\epsilon}$ and its components are expressed by
\begin{eqnarray}
\bm{\epsilon} &=& \bm{\epsilon}_e  + \bm{\epsilon}_{t} + \bm{\epsilon}_p, \label{eq_fea4}\\
\bm{\epsilon}_e &=& \frac{1+\nu}{E}\bm{\sigma} - \frac{\nu}{E} \Tr(\bm{\sigma})\mathbb{I}, \label{eq_fea5}\\
\bm{\epsilon}_{t} &=& \left[\epsilon_{t} \, \epsilon_{t} \, \epsilon_{t} \, 0 \, 0 \, 0 \, \right]^{\text{T}}, \\
\bm{\epsilon}_{t} &=& \alpha(T-T_0), \label{eq_fea7} \\
\bm{\epsilon}_p &=& g(\sigma_Y), \label{eq_fea6}\\
f &=& \sigma_m - \sigma_y(\bm{\epsilon}_q, T) \leq 0,  \\
\bm{\dot{\epsilon}}_p &=& \bm{\dot{\epsilon}}_q \bm{a}, \\
\bm{a} &=& \left( \frac{\partial f}{\partial \bm{\sigma}} \right)^{\text{T}}, \\
\end{eqnarray}
where $E$ is the Young’s modulus, $\nu$ the Poisson’s ratio, $\mathbb{I}$ the identity matrix, $\alpha$ the coefficient of thermal expansion, $T$ the current nodal temperature, $T_0$ the initial reference temperature, $g(\sigma_Y)$ a function related to the yield strength $\sigma_Y$, $f$ the yield function, $\sigma_m$ the Mises' stress, $\sigma_y$ the initial yield strength, $\bm{\epsilon}_q$ the equivalent plastic strain, and $\bm{a}$ the flow vector. $\dot{\epsilon}$ denotes the time derivative $\frac{\partial \epsilon}{\partial t}$. It should be noted that total strain $\bm{\epsilon} \ll 1$ assumes that the deformation is small, which is the case in the LPBF process. In this study, a plastic behavior with isotropic strain hardening is assumed and hence, the plastic strain $\bm{\epsilon}_p$ is computed by enforcing the von Mises yield criterion and the Prandtl-Reuss flow rule. The stress-strain behavior beyond the yield point is described by the tangent modulus as specified by the ANSYS BISO metal plasticity model (see below). Essentially, this model describes the function $g(\sigma_Y)$ in Eq.~(\ref{eq_fea6}). As mentioned earlier, the Newton-Raphson method employs multiple iterations until force equilibrium is achieved. 

At the beginning of structural calculations, the heat transfer FEA model is automatically converted from a model comprising thermal SOLID 70 elements to that with SOLID 145 structural elements. Temperature dependent mechanical properties (e.g., $E$, $\nu$, $\sigma_y$, $\alpha$) for the Ti-6Al-4V alloy are presented in Table~\ref{table_thermophysical}.
\begin{table}[h]
\begin{center}
\caption{Temperature dependent thermophysical and mechanical properties used in FEA simulation, after ~\cite{Yang2015}.}\label{table_thermophysical}
  \begin{tabular}{| c | c | c | c | c | c |}
    \hline
    $T$ ($^\circ$C) & $E$ (GPa) & $\sigma_Y$ (MPa) & $\alpha$ (K$^{-1}$) & $\nu$ & Tangent Modulus (MPa) \\ \hline
    25 & 105.863 & 879.997 & \num{8.8480e-6} & 0.2985 & 1250.0 \\ \hline
   200 & 95.976 & 850.612 & \num{9.9458e-6} & 0.3041 & 1125.0 \\ \hline
   400 & 84.676 & 680.892 & \num{9.9535e-6} & 0.3105 & 1000.0 \\ \hline
   600 & 73.376 & 511.172 & \num{1.0234e-5} & 0.3169 & 222.2 \\ \hline
   800 & 62.076 & 341.452 & \num{1.0299e-5} & 0.3233 & 111.0 \\ \hline
   1000 & 50.776 & 171.732 & \num{1.0149e-5} & 0.3297 &  \\ \hline
   1100 & 45.126 & 86.872 & \num{1.0291e-5} & 0.3329 &  \\ \hline
   1200 & 39.476 & 50.860 & \num{1.0291e-5} & 0.3361 &  \\ \hline
   1400 & 28.176 & 14.860 & \num{1.0291e-5} & 0.3425 &  \\ \hline
   1600 & 16.876 & 14.860 & \num{1.0291e-5} & 0.3489 &  \\ \hline
   1800 & 5.576 & 14.860 & \num{1.0291e-5} & 0.3553 &  \\ \hline
   2000 & 5.576 & 14.860 & \num{1.0291e-5} & 0.3617 &  \\ \hline
  \end{tabular}
  \end{center}
  \end{table}
  Material models for plasticity are considered using ANSYS BISO model (Bilinear isotropic hardening plasticity)~\cite{Ansys}. This is the classical bilinear isotropic hardening model which utilizes two slopes (elastic and plastic) to represent the stress-strain behavior of a material. The temperature dependent yield stress and tangent moduli data are taken from Ref.~\cite{Yang2015}. This BISO model uses von Mises yield criteria with an isotropic hardening assumption. The initial slope of the curve is the Young’s modulus. At the specified yield stress, the stress-strain curve continues along the second slope defined by the tangent modulus. 

The stress analysis can be conducted using temperature data from any step of the corresponding thermal simulation using the computed temperature distribution in the model at that time. In the present study, the structural analysis is conducted when the laser has traversed approximately 80\% of the entire length. Such analyses were run for all five laser power levels used in the study. A typical thermal profile for the case of the 20 W laser is shown in Fig.~\ref{figure_fea}. The structural analysis uses the computed temperature field, temperature-dependent mechanical properties, and phase proportions at the end of a particular step of the heat transfer analysis to compute the mechanical behavior. The nonlinear mechanical analysis is treated as a quasi-static incremental analysis. The total strain comprises elastic, plastic, and thermal strains. The model is physically constrained by providing fixed boundary conditions to the bottom nodes of the base plate. 

\begin{figure}
\begin{center}
\includegraphics[scale=0.75]{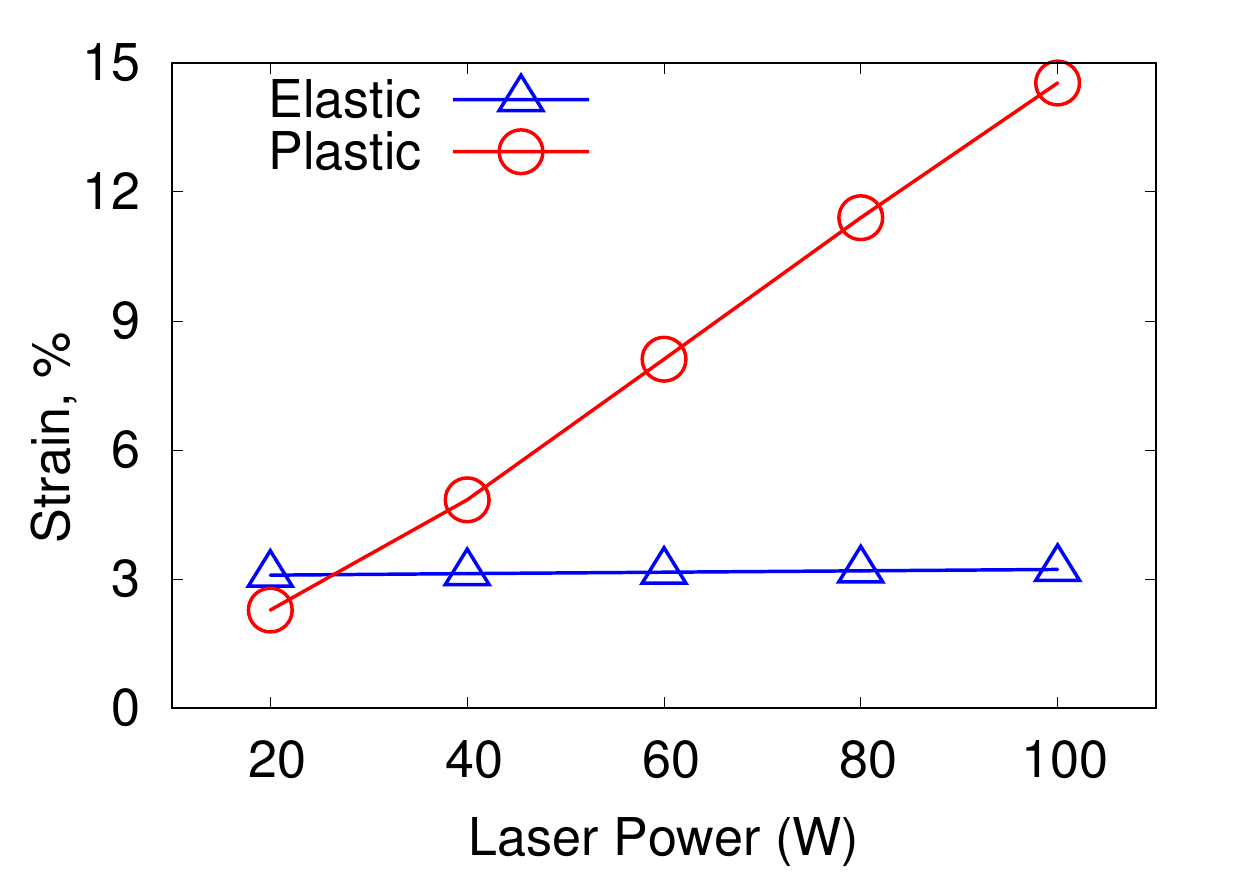}
\caption{The maximum value of the von Mises effective elastic strain remains constant and the maximum value of the plastic strain increases with increasing value of the laser power. Figure~\ref{figure_fea} shows the location of the laser center when the laser has traversed 80\% of the total length of 2.5 mm. While the maximum value of the von Mises elastic strain is obtained almost at the laser center, the maximum von Mises plastic strain is obtained at the perimeter of the laser center.}\label{figure_strain}
\end{center}
\end{figure}

\begin{figure}
\begin{center}
\includegraphics[scale=0.3]{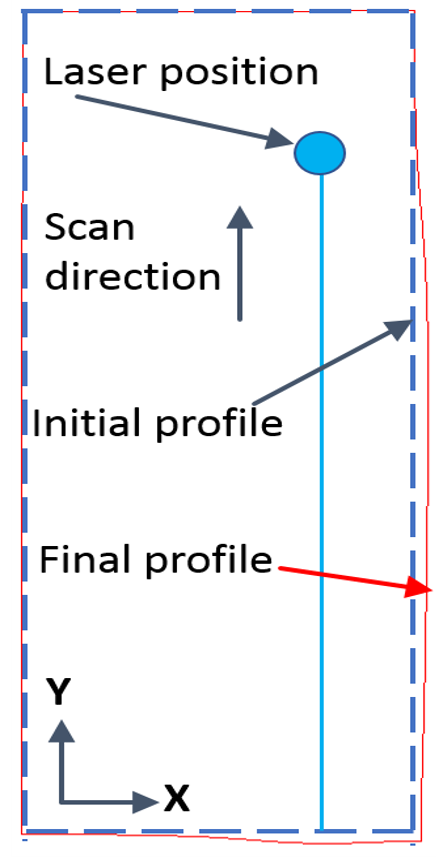}
\caption{A plan view of the part at mid depth of the powder bed showing distortion for the case of 60 W laser after stress analysis for the laser position (0.25 mm, 2 mm, 0 mm) shown. On the projection of this plane, the red edges show the final boundary of the plane indicating distortion in the top surface (not to scale).}\label{figure_distortion}
\end{center}
\end{figure}

\begin{figure}
\centering
\subfloat[]{\includegraphics[scale=0.25]{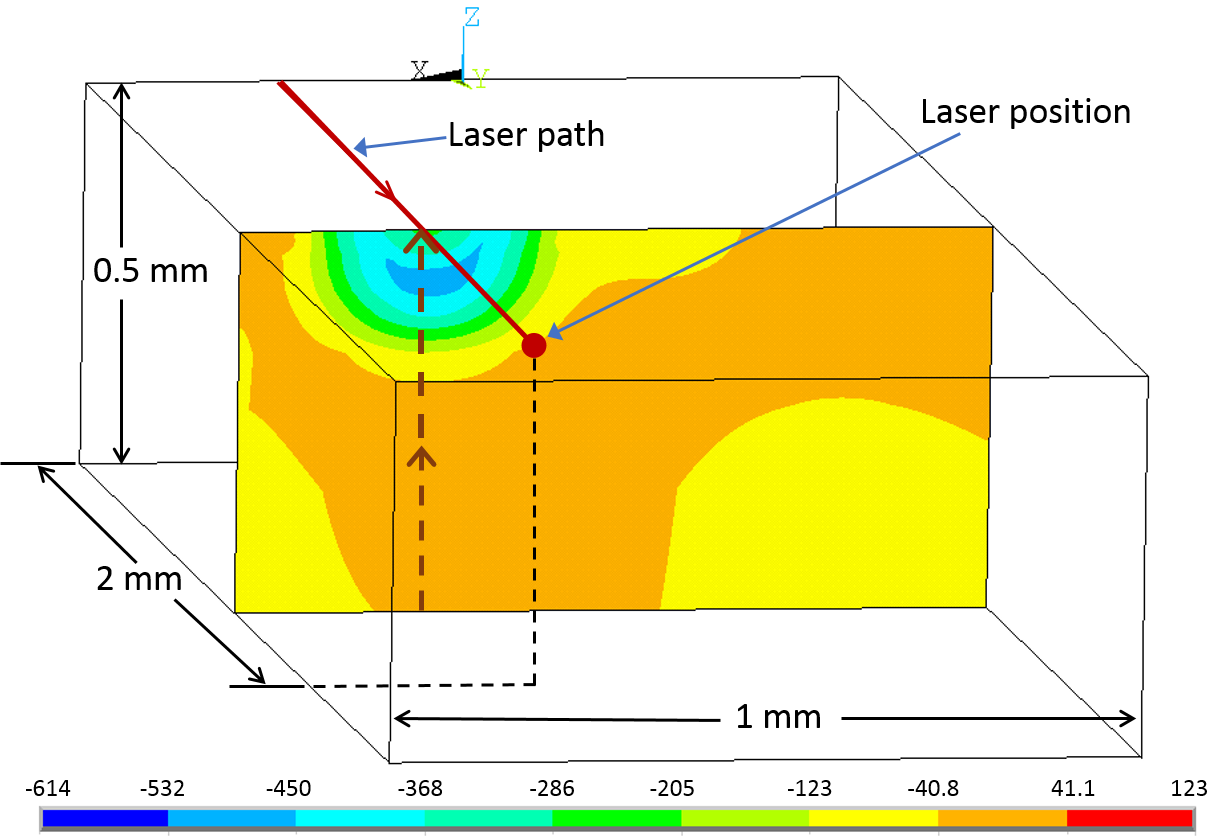}}\hspace{-5mm}\hfill
\subfloat[]{\includegraphics[scale=0.4]{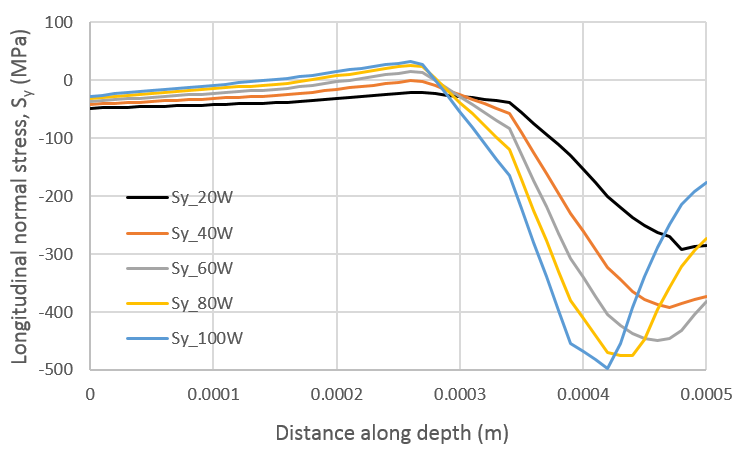}}\hspace{-5mm}\hfill
\caption{(a) A plot of residual normal stress, $S_y$, on a transverse cut plane at mid-length of the part for the case of the 80 W laser. Stress units are in MPa. (b) Variation of $S_y$ along the depth (at mid length of the part from the bottom to the top surface directly below the position of the laser) is estimated along the arrow in Fig.~\ref{figure_stress}a and is plotted for various values of laser power.}\label{figure_stress}
\label{figure}
\vspace{-4mm}
\end{figure}

It is expected that the overall elastic and plastic strains increase with increase in laser power (Fig.~\ref{figure_strain}). As expected, the elastic strain is very small. However, a large plastic strain of 14.5\% was reached for the 100 W laser. As the laser power increases, the von Mises effective stress increases (not shown), which is expected since a large laser power has more drastic effect on the temperature distribution. The complex thermal profile and the mechanical restraint result in both tensile and compressive stresses in the part. Such a distribution in the residual tensile and compressive stresses provides insight into how the part will behave when put to service. The stress analysis also provides a description of distortion in the part as a result of the deposition process. A typical plot (plan view) of the model (at mid-depth of the powder bed) at the end of stress analysis is shown in Fig.~\ref{figure_distortion} (before the part has cooled to room temperature) showing distortions. As evident in this plot, there is no thermal distortion in the region ahead of the laser front. Structural analysis also provides valuable insights into the residual stress distribution in the formed part. Figure~\ref{figure_stress}a shows the longitudinal normal stress ($S_y$) distribution on a transverse plane at the mid-length of the part at the instant when the 80 W laser has traversed 80\% of the length of the part. It is evident that large compressive residual stresses develop just below the melt pool region. Figure~\ref{figure_stress}b shows $S_y$ profiles from bottom to top along the depth at regions just below the laser (at mid-length of the part). Largest compressive stresses develop for the case of the 100 W laser, as expected. The ratio of the maximum compressive stress to the baseline stress varies from 6.1 for the 20 W laser to 18.2 for the 100 W laser. Also, the maximum compressive stress point shifts downward as the laser power is increased, which is expected. Regions far away from the laser show very little residual stress. This is true for regions at the bottom of the substrate. The residual stress profile obtained here matches very well with those in Ref.~\cite{Beuth2003}, where a compressive residual stress was reported below the melt pool and mildly positive (tensile) stress was seen at the vertical free end. Note that in Ref.~\cite{Beuth2003} traction free boundary conditions were used at the vertical free end of the part, which were not used here. It is clear that laser power should be kept at a minimum level if large residual stresses and distortions are to be avoided.


\begin{figure}
\begin{center}
\includegraphics[scale=0.7]{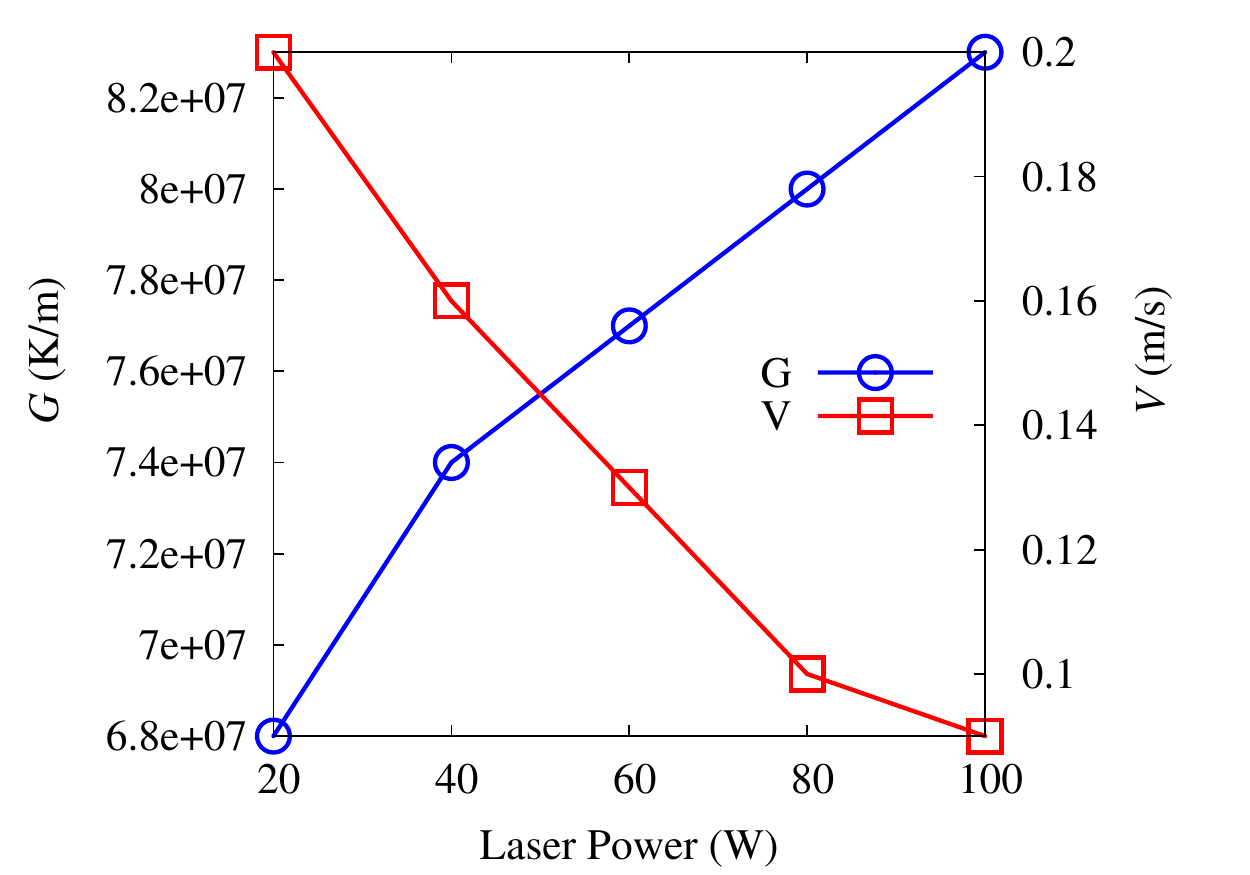}
\caption{Average temperature gradient $G$ and solidification velocity $V$ are estimated from the solidification front of the melt pool in Fig.~\ref{figure_fea} for various values of laser power.}\label{figure_fields}
\end{center}
\end{figure}

The FEA simulated temperature distribution and mechanical analysis can be correlated with the solidification process through the calculation of temperature gradient and solid-liquid interface velocity for different values of laser power. The trailing edge of the melt pool is the solidification front, where prior-$\beta$ phase nucleates and grows roughly perpendicular to the solid-liquid interface at a velocity $V$ in the direction of maximum temperature gradient $G$. $G$ in the melt pool is estimated by $(T_P - T_L)/r$, where $T_P$ is the maximum temperature in the melt pool, $T_L$ = 1928 K is the equilibrium liquidus temperature, and $r$ is the distance from the point along the solidification boundary to the point of maximum temperature in the melt pool. $V$ is estimated by $V_b \cos\alpha$, where $V_b$ is the beam speed and $\alpha$ is the solidification angle between the normal to the fusion boundary and the laser scanning direction. In our simulations, the average $G$ along the solidification front increases from 6.8 $\times$ 10$^7$ K m$^{-1}$ to 8.3 $\times$ 10$^7$ K m$^{-1}$ and the average $V$ decreases from 0.2 m s$^{-1}$ to 0.1 m s$^{-1}$ for a fixed laser scan speed of 800 mm s$^{-1}$ and increasing values of laser power from 20 W to 100 W (Fig.~\ref{figure_fields}).
Since none of the calculated ranges for the solidification conditions are high, the net effects of either $G$ or $V$ on the microstructural evolution through the cooling rate ($GV$) are equivalent. We use a constant $V$ = 0.15 m s$^{-1}$ in our simulations to estimate the microstructural fields as a function of $G$. Note that during solidification of Inconel 718, on average, $G$ varied between 0.14 $\times$ 10$^7$ K m$^{-1}$ and 2.4 $\times$ 10$^7$ K m$^{-1}$ and $V$ varied between 0.01 m s$^{-1}$ and 0.3 m s$^{-1}$~\cite{Trevor2017}. These differences are due to thermal diffusivity and density differences between these alloys~\cite{Debroy2017}, leading to significant differences in $T_P/T_L$ values in the melt pool. $T_P/T_L$  $\approx$ 12 for Ti-6Al-4V, while $T_P/T_L$  $\approx$ 2 for Inconel 718 for $P$ = 100 W and $V_b$ = 800 mm s$^{-1}$.

\section{Phase-field simulation}
We use the above $G$ and $V$ values in a quantitative binary alloy phase-field model detailed in Ref.~\cite{Echebarria2004} to simulate the melt pool solidification process of a pseudo-binary Ti alloy melt into prior-$\beta$ phase. The model uses the following time dependent partial differential equations for the phase field ($\phi$) and concentration ($c$) field variables, which are solved iteratively to obtain the steady state microstructure fields. The evolution equation for $\phi$ is given by
\begin{eqnarray}\label{eq_phi}
\tau_0 a(\hat{n})^2\frac{\partial \phi}{\partial t} = W_{0}^{2} \nabla \cdot \left[{a(\hat{n})}^2 \nabla\phi\right] + \sum_{i=1}^{d} \partial_i \left[a(\hat{n}) \frac{\partial a(\hat{n})}{\partial(\partial_i \phi)} |\nabla\phi |^2 \right] \nonumber \\
+ \phi -\phi^3 -  \frac{\lambda}{1-k_e} (1-\phi^2)^2 \left[\exp(u) -1 + \frac{G(z-Vt)}{m_l c_0/k_e}\right].
\end{eqnarray}
$a(\hat{n}) = 1 - \epsilon_4 \left[3- 4 \sum_{i=1}^{d} n_{i}^{4}\right]$ represents the solid-liquid interface energy with magnitude $\epsilon_4$, and $n_i$ is the interface normal vector pointing into liquid along the Cartesian directions $i$ in dimensions $d$. The thermophysical properties data including alloy composition $c_0$, liquidus slope $m_l$, and equilibrium partition coefficient $k_e$ are given in Table~\ref{table_param_pf}. $u$ is the dimensionless chemical potential given by $\ln \left(\frac{2ck_e/c_0}{1+k_e-(1-k_e)\phi}\right)$. The temperature gradient $G$ is translated along the $z$ (growth) axis with a velocity $V$ in a directional frozen-temperature framework. The evolution equation for $c$ is given by
\begin{equation}\label{eq_c}
\frac{\partial c}{\partial t} = -\nabla \cdot \left[ - \frac{1}{2}(1+\phi)\, D_l \, c \, \exp(u)^{-1}  \, \nabla\exp(u) +\frac{1}{2\sqrt{2}} W_0 (1-k_e) \exp(u) \frac{\partial \phi}{\partial t} \frac{\nabla\phi}{|\nabla \phi|}\right],
\end{equation}
where the first term inside the square bracket represents a standard Fickian diffusion flux and the second term is the anti-trapping solute flux which was introduced~\cite{Echebarria2004} to avoid unphysical solute-trapping effects due to the use of large numerical interface thickness values in simulations leading to effective solute distribution at solid-liquid interfaces. $D_l$ is the diffusivity of solute in the liquid.

\begin{table}[h]
\caption{Thermophysical properties used in phase-field simulations, taken directly after~\cite{Chou2016,nastac2012}.}\label{table_param_pf}
\centering
\begin{tabular}{ll}
\hline
Alloy Mass Fraction ($c_0$) &\SI{10}{\%} \\
Equilibrium Partition Coefficient ($k_e$)	&0.838	\\
Liquidus Slope ($m_l$) &-0.088 K \%$^{-1}$	\\
Equilibrium Freezing Range ($\Delta T_0$) &50 K	\\
Liquid Diffusion Coefficient ($D_l$) &$9.5 \times 10^{-9}$ m$^2$ s$^{-1}$ \\
Anisotropy Strength ($\epsilon_4$) &\SI{3}{\%}	\\
Gibbs-Thomson coefficient ($\Gamma$) &$2.2 \times 10^{-7}$ K m \\
\hline
\end{tabular}
\end{table}%

The numerical parameters in this model -- the interface thickness ($W_0$), the phase-field relaxation time ($\tau_0$), and the coupling constant ($\lambda$) -- are linked to the thermophysical properties via the chemical capillary length $d_0 = a_1 W_{0}/\lambda$ and the time scale for diffusion $\tau_0 = a_2\lambda W_{0}^{2}/D_l$ using a thin-interface analysis which makes the interface kinetics vanish. The Gibbs-Thomson constant is given by $d_0$ times the freezing range of the alloy (Table~\ref{table_param_pf}). The numerical constants are given by $a_1$ = 0.8839 and $a_2$ = 0.6267, after Ref.~\cite{Echebarria2004}. Both $W_0$ and $\tau_0$ values are used to render all the simulation parameters dimensionless. Note that the present model is based on significant simplifications of LPBF experiments and ignores convection in the liquid, diffusion of heat and solid-liquid interface kinetics (which is set by the constants $a_1$ and $a_2$). Therefore, present simulations can be considered as a baseline reference for LPBF microstructure evolution during the solidification of a dilute pseudo-binary Ti-\SI{10}{\%}X melt (approximating Ti-6Al-4V). This is a standard procedure for the quantitative phase-field solidification model~\cite{Echebarria2004,Karma2001} used here.

\begin{figure}
\begin{center}
\includegraphics[scale=0.5]{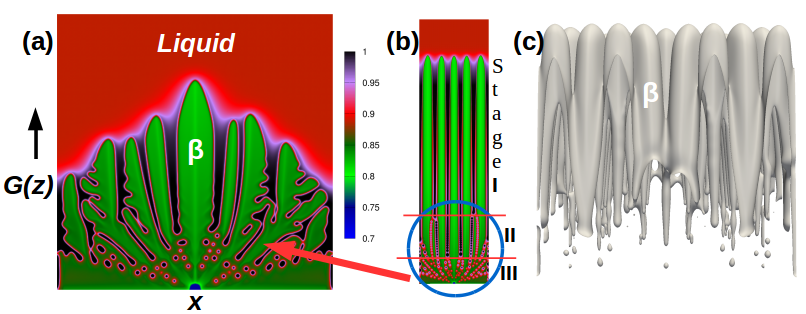}
\caption{(a) Transient state morphology of as-solidified $\beta$(Ti) is presented for $G$ = 7.4 $\times$ 10$^7$ K m$^{-1}$ and $V$ = 0.15 m s$^{-1}$. Onset of columnar morphology with secondary arms is seen in early stage. Solute concentration profile (scaled by $c_0/k_e$) is shown using a color scheme. (b) Full columnar $\beta$-phase results in the steady state with a constant $\beta$ interspacing $\lambda_\beta$. Growth stages I, II and III are related to Fig.~\ref{figure_coarsening}. (c) Full columnar morphology forms in three dimensions. A few solute-enriched droplets pinch off from the cell roots.}\label{figure_pf}
\end{center}
\end{figure}

Equations~(\ref{eq_phi}) and~(\ref{eq_c}) are solved on a uniform mesh using a finite volume method and no-flux boundary conditions in all directions. The numerical values of the parameters used in the simulations are: interface thickness of 5 nm; grid spacing of 4 nm; dimensionless timestep ($t/\tau_0$) of 0.02; and mesh size (in grid units) of 600$\times$3000 in 2D and 240$\times$240$\times$600 in 3D. The interface thickness and grid spacing used in our simulations are small enough that the results become independent of their values. Simulations begin with an initial $\beta$(Ti) circular/spherical seed of radius \SI{0.04}{\micro\meter} at the bottom of the simulation box, which initially grows perpendicular to the solid-melt interface (Fig.~\ref{figure_pf}a) and eventually grows in the (vertical) build direction with a velocity $V$ and temperature gradient $G$ estimated by FEA simulations to reach a fully columnar microstructure in the steady state (refer to Figs.~\ref{figure_pf}b and~\ref{figure_pf}c). 

\subsection{Microstructure phase fields}
\begin{figure}
\begin{center}
\includegraphics[scale=0.75]{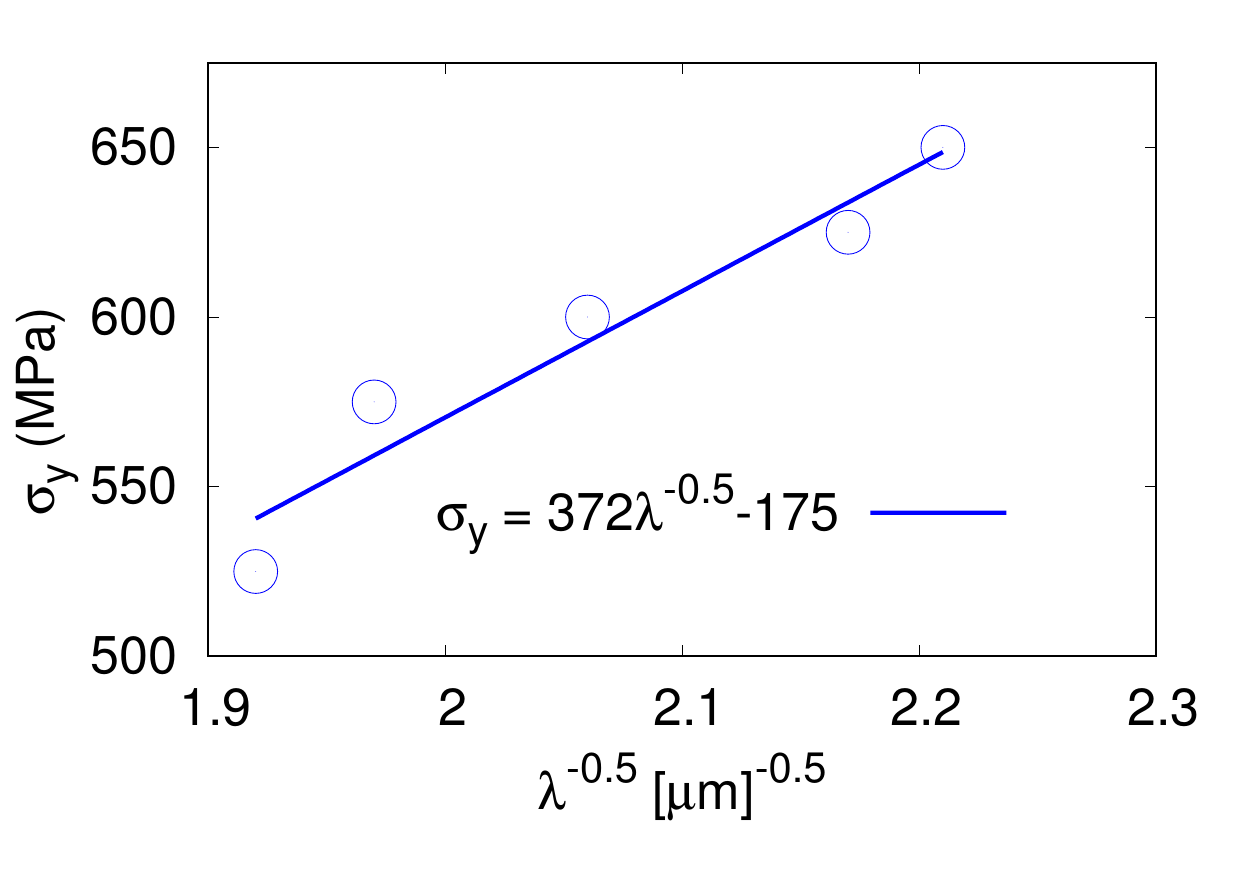}
\caption{Correlation of $\lambda_\beta$ with the von-Mises effective stress follows the Hall-Petch relation. For details, refer to text.}\label{figure_spacing}
\end{center}
\end{figure}

The $\beta$ phase field (from 2D simulations) is plotted in Fig.~\ref{figure_pf}b to reveal the fully columnar microstructure in steady state. Secondary sidebranches appear in the initial transient stages of growth, as shown in Fig.~\ref{figure_pf}a, which are absent in steady state. The corresponding concentration field illustrates the diffusion field around the $\beta$ phase and the spatial distribution (or microsegregation) of solute in the melt within the columnar $\beta$ network. The size of $\beta$ phase varies with the cooling rate, or precisely by $G$ in our simulations. The steady state $\beta$ interspacing $\lambda_\beta$ (in Fig.~\ref{figure_pf}b) is measured using the Fourier transform of the solid-liquid interface profile in a power spectrum analysis, as described in Ref.~\cite{Greenwood2004,supriyo20173d}. The microstructural scale $\lambda_\beta$ is significant since it determines the yield strength of as-deposited Ti-6Al-4V. In the present work, yield strength ($\sigma_y$) is taken as the von Mises effective stress predicted by FEA simulations. Therefore, $\lambda_\beta$ can be correlated with the yield strength following the Hall-Petch relationship given by $\sigma_y = \sigma_0 + K\lambda_{\beta}^{-0.5}$~\cite{Tan2015}, where $\sigma_0 = -175$ MPa and $K = 372$ MPa $[{\SI{}{\micro\meter}}]^{0.5}$ are material constants. This is illustrated in Fig.~\ref{figure_spacing}, where the yield strength of as-deposited $\beta$(Ti) phase increases with decreasing microstructural scale ($\lambda_\beta$) which corresponds to increasing laser power. The size of the $\beta$ phase is extremely fine in our simulations, ranging between \SI{0.2}{\micro\meter} and \SI{0.3}{\micro\meter} and the predicted yield stress for these $\lambda_\beta$ values ranges between 500 MPa and 650 MPa. These $\sigma_y$ values are consistent with previous studies~\cite{Formanoir2016,Gorsse2017,Tan2015}.

\begin{figure}
\begin{center}
\includegraphics[scale=0.75]{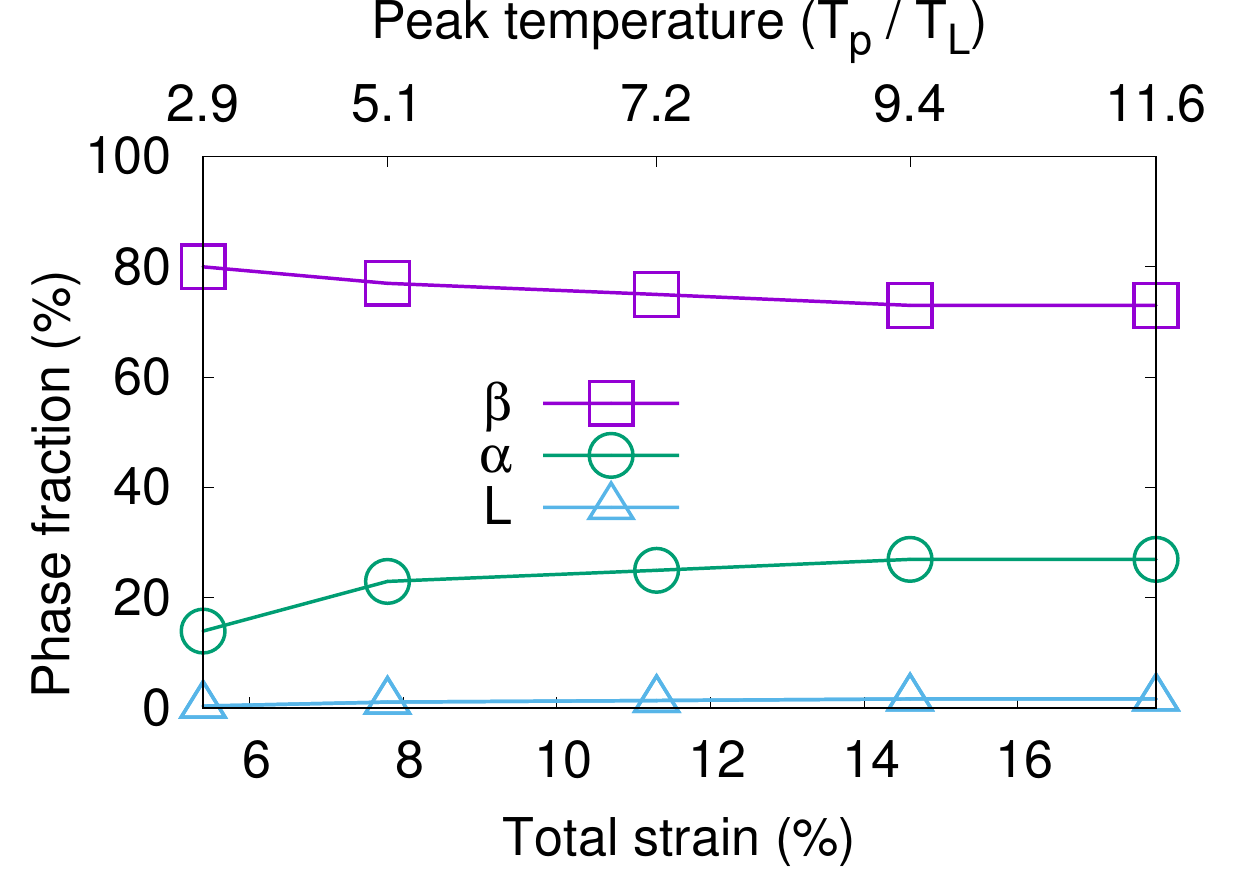}
\caption{Phase fraction of $\beta$, $\alpha$ and $L$ phases are estimated after melt solidification and are correlated with total strain estimated from FEA simulations for various heat input values. Total strain increases as the amount of prior-$\beta$ decreases with increasing heat input or increasing ratio of maximum temperature ($T_P$) in the melt pool to the equilibrium liquidus temperature ($T_L$). Consequently, the amount of $\alpha$ increases and the fraction of $L$ phase that could potentially form along the $\alpha/\beta$ interface is estimated small.}\label{figure_volume}
\end{center}
\end{figure}

The distribution of microstructure phases is important for tensile and fatigue properties. Since we work on a binary alloy, we model only $\beta$(Ti) and liquid phases. The phase fraction of $\beta$ we obtain from phase-field simulations is the maximum amount of $\beta$ one would expect in the final microstructure at room temperature (Fig.~\ref{figure_volume}). And the amount of residual liquid in between $\beta$ columns can be considered as the minimum amount of $\alpha$ in the final microstructure. The prior $\beta$ phase will continue to transform to $\alpha$ phase with increasing undercooling below the $\beta$ transus temperature during the heat treatment in solid state, which we could not capture in the present model. Eventually, at room temperature, the microstructure will consist of primarily $\alpha$ phase in basket-weave and/or colony structures, which are often separated by prior $\beta$ grain boundaries~\cite{Tan2015,Lore2010}. The optimum phase fractions in the final microstructure for desired material properties can be controlled by the solidification process. We correlate the effective strain (elastic + plastic) estimated from FEA simulations with the phase fractions estimated from phase-field simulations for reference. $\beta$ phase is relatively softer compared to the $\alpha$ phase. The effective strain increases from $\approx$ 5\% to 18\% for increasing values of the laser power during FEA mechanical analysis, the $\beta$ fraction decreases from 80\% to 73\% in our phase-field simulations for the solidification conditions corresponding to the laser power values used in FEA thermal analysis (Fig.~\ref{figure_volume}). The intermediate $L$ phase forms during $\beta \leftrightarrow \alpha$ transition along the $\alpha$/$\beta$ interface under certain conditions during experiments~\cite{Tan2015}. We take the interface contour area of three-dimensional phase fields (in Fig.~\ref{figure_pf}c) as an estimation of $L$ fraction (Fig.~\ref{figure_volume}). The estimated $L$ fraction is below 1\% and increases with increasing heat input. This is in agreement with previous experiments~\cite{Tan2015} where $L$ fraction was found to be very small and could not be detected by conventional XRD techniques. With increasing heat input during an additive manufacturing process, the segregation of the solute element increases, which in turn promotes the formation of $L$ phase. Similar observations were reported in experiments~\cite{Tan2015}. Simulations with multiple phase-fields~\cite{Tamoghna2017} would further help to understand its formation and growth mechanisms in our future work.

\begin{figure}
\begin{center}
\includegraphics[scale=0.75]{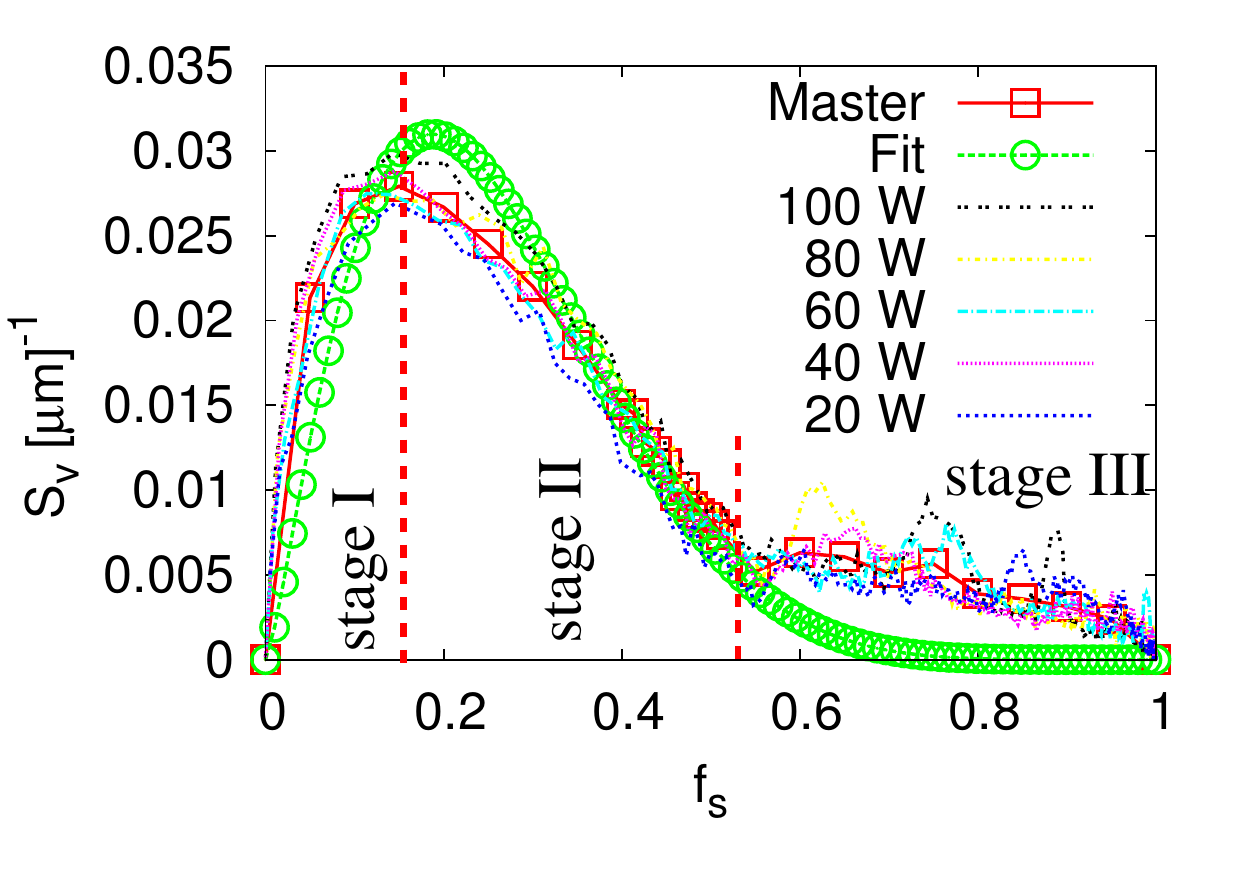}
\caption{The interfacial area density $S_v$ varies with the solid fraction $f_s$, the behavior of which is explained in the text.}\label{figure_coarsening}
\end{center}
\end{figure}

Next, we calculate the interfacial surface area density $S_v$ over the solidification distance from our simulated three-dimensional cells in Fig.~\ref{figure_pf}c. This information is critical to the formation of solidification defects. In experiments, $S_v$ is estimated from the metallographic sections of X-ray tomography solidification data~\cite{Gibbs2015}. For each $x-y$ horizontal section in our 3D simulations, $S_v$ is calculated by $f_sA/V_s$, where $f_s$, $A$ and $V_s$ all are determined from that particular section; $f_s$ is the fraction of the solid, $A$ the length of the interface approximated by the discrete number of interface points and $V_s$ the area contained by the solid $\beta$ phase. $S_v$ is generally expressed as $f_{s}^{p} (1-f_s)^q$~\cite{Neumann2017}, where $p$ and $q$ are constants due to a particular process that involves only growth with no curvature driven coarsening, as in our present case. The variation of $S_v$ with $f_s$ is shown in Fig.~\ref{figure_coarsening}. $f_s = 0$ is close to the top and $f_s = 1$ is close to the bottom of the simulation box. $S_v$ increases with $f_s$ (stage I: late stage) due to free cellular growth, creating new interfaces over time. Following a maximum signifying the onset of coalescence of the cell-liquid interfaces in the semisolid mushy zone at $f_s = 0.2$, $S_v$ decreases (stage II: intermediate stage) due to impingement and coalescence of the interfaces. Note that the $S_v$ data in the solid (stage III: early stage) is noisy due to complex solute interaction and rapid interface coalescence processes. Therefore, we construct a master $f_s$ vs. $S_v$ curve using the mean of the $S_v$ values obtained from each case, and we found the master curve to fit best with the expression: $S_v = f_{s}^{1.4} (1-f_s)^{5.8}$. The values of $p$ and $q$ are not available in the literature for rapid solidification conditions in the high $G$, high $V$ limit. In the low $G$, low $V$ solidification limit, the exponents $p$ and $q$ varied between 0.5 and 1 in the literature, depending on the solidification conditions and the geometry of the evolving interfaces~\cite{Neumann2017,Ratke2010}. In the low limit of $G$ and $V$, both $p$ and $q$ were found to decrease in Ref.~\cite{Neumann2017} and the cell-liquid interfaces coalesced at a higher solid fraction $f_s > 0.5$. In our simulations, coalescence begins at a smaller solid fraction ($\approx$ 0.2), thus more liquid is retained in the mush, making the $\beta$(Ti) columns vulnerable to porosity and other solidification defects. 

\section{Summary and Outlook}
In summary, we performed finite element simulations to predict the temperature and stress fields and phase-field simulations to predict the $\beta$ microstructure fields during laser powder bed fusion of Ti-6Al-4V, using various values of laser power between 20 W and 100 W and fixed laser scan speed of 800 mm s$^{-1}$. The heat transfer analysis indicates that the average $G$ along the solidification front increases from 6.8 $\times$ 10$^7$~ K~m$^{-1}$ to 8.3 $\times$ 10$^7$~K~m$^{-1}$ and the average $V$ decreases from 0.2 m s$^{-1}$ to 0.1 m s$^{-1}$ as the laser power increases from 20 W to 100 W. The stress analysis of the deposited part indicates that the resultant elastic strain is very small, while the plastic strain significantly increases with increasing laser power and varies between 2.5\% and 14.5\% in our simulations. The stress analysis also determines the residual stress distribution in the formed Ti-6Al-4V part, where large compressive residual stresses develop just below the solidified melt pool region, the magnitude of which increases with increasing values of the laser power. The ratio of the maximum compressive stress to the baseline stress varies from 6.1 for the 20 W laser to 18.2 for the 100 W laser.

Phase-field simulations use the $G$ and $V$ data for the simulation of $\beta$(Ti) columnar microstructures. The columnar spacings are extremely fine in our simulations, ranging between $\approx$ \SI{0.2}{\micro \meter} and \SI{0.3}{\micro \meter}, which are correlated with the von Mises effective stress values ranging between 525 MPa and 650 MPa estimated from FEA simulations using a Hall-Petch equation. The post solidification phase fraction of the resulting $\beta$ field varies between 73\% and 80\% depending on the heat input in the simulations and is correlated with the total effective strain values estimated from FEA simulations. We express the interfacial surface area density as a function of $f_{s}^{1.4} (1-f_s)^{5.8}$ where coalescence of the $\beta$(Ti) columns begins at a smaller $\beta$ fraction $f_s$ $\approx$ 0.2, leading to poor mechanical behavior of the as-deposited Ti-6Al-4V part in the rapid solidification limit.

Although the prior-$\beta$ phase fraction decreases after the solution heat treatment in solid state, the full columnar morphology of $\beta$ is retained~\cite{Vastola2016,Korner2016}. This is precisely due to rapid cooling rates that result during a laser based additive manufacturing process, while the columnar $\beta$ morphology is modified significantly due to lower cooling rates during an electron beam melting process~\cite{Vastola2016,Korner2016}. We always obtain a fully columnar $\beta$(Ti) microstructure for the solidification conditions simulated in the present work. This is in complete agreement with the previously determined~\cite{Kobryn2001} solidification (or $G-V$) map for cast Ti-6Al-4V. The volume fraction of solute-rich droplets that pinch off from the $\beta$ roots (Fig.~\ref{figure_pf}c) is far smaller compared to nickel alloys~\cite{supriyo2017,ghosh2018_droplet}, signifying a weak microsegregation in Ti alloys and hence absence of secondary solid phases. The effects of convection in FEA simulations are less pronounced since the thermal diffusivity of Ti alloys is large~\cite{Debroy2017} and are ignored as a first approximation. The effects of convection on the $\beta$ interspacing are not as pronounced as compared to the secondary arms~\cite{Lee2010}, which are absent in steady state. The simulated $\beta$ cells are extremely fine and provide significant resistance to fluid flow following an exponential increase of the damping effect in the mushy region, leading to reduced effects of convection~\cite{Tan2011}. This is why computational fluid dynamics analysis is not considered in this study. Therefore, simulations have been performed with reasonable approximations to predict the temperature, stress and microstructure fields in Ti-6Al-4V. 

Validation of our numerical models can be obtained through the comparison of our simulation results with experimental observations. We are not aware of any comprehensive experimental study of coupled thermo-mechanical and microstructure analyses of LPBF of Ti-6Al-4V. Although we do not perform experimental validation of our microstructure and mechanical analyses, our phase-field simulated solidification morphology and grain size distribution qualitatively agree with the experiments reported in Ref.~\cite{Tan2015}, and the yield strength values approximated from our finite element simulations are reasonable with the experiments reported in Refs.~\cite{Formanoir2016,Gorsse2017,Tan2015}. Motivated by our present simulations, we plan on experimental validation in the future. In addition, a multi-component phase-field framework~\cite{Ghosh2017_eutectic} will be used to represent the $\beta$, liquid and $\alpha$ phases in the microstructure, and melt convection will be considered for more accurate microstructure-property correlation.

\section*{References}
\bibliography{papers}

\end{document}